\RequirePackage[l2tabu,orthodox]{nag} 
\documentclass[12pt,letterpaper]{article} 

\usepackage[T1]{fontenc} 
\usepackage[utf8]{inputenc} 
\usepackage{ebgaramond} 
\usepackage{tgheros} 
\usepackage[USenglish]{babel} 
\usepackage[strict,autostyle]{csquotes} 
\usepackage[babel=true]{microtype} 

\usepackage{amsmath} 
\usepackage{authblk} 
\usepackage{booktabs} 
\usepackage{multirow} 
\usepackage{caption} 
\usepackage{datetime} 
\usepackage[final]{draftwatermark} 
\usepackage{endnotes} 
\usepackage{geometry} 
\usepackage{graphicx} 
\usepackage{natbib} 
\usepackage{rotating} 
\usepackage{setspace} 
\usepackage{titlesec} 
\usepackage{url} 

\usepackage{hyperref} 
\usepackage{orcidlink} 

\usepackage[usestackEOL]{stackengine}
\usepackage{caption}
\usepackage{multirow}
\usepackage{graphicx}
\usepackage{tabularx}
\usepackage{makecell}

\newdateformat{monthyeardate}{\monthname[\THEMONTH] \THEYEAR}

\newcommand{\myname}{Geoff Boeing, Jiyoon Kim, Daniele Simon-Thompson, Jaehyun Ha, Ruoyu Chen}
\newcommand{\myaffiliation}{Department of Urban Planning and Spatial Analysis\\University of Southern California}
\newcommand{\paperdate}{\monthyeardate\today}
\newcommand{\papertitle}{Choose Your Own Adventure: How Data Source Selection Shapes Urban Accessibility Analyses}
\newcommand{\papercitation}{Boeing, G., J. Kim, D. Simon-Thompson, J. Ha, and R. Chen. 2026. \papertitle. \textit{Journal of Transport Geography}, published online ahead of print. \href{https://doi.org/10.1016/j.jtrangeo.2026.104841}{doi:10.1016/j.jtrangeo.2026.104841}}
\newcommand{\paperkeywords}{Accessibility, Urban Geography, Parks, Urban Planning, Transportation, Data Science}

\graphicspath{{.}}

\titleformat{\section}{\normalfont\sffamily\large\bfseries\color{black}}{\thesection.}{0.3em}{}
\titleformat{\subsection}{\normalfont\sffamily\small\bfseries\color{black}}{\thesubsection.}{0.3em}{}
\titleformat{\subsubsection}{\normalfont\sffamily\small\color{black}}{\thesubsubsection.}{0.3em}{}

\hypersetup{
  pdfauthor={\myname},
  pdftitle={\papertitle},
  pdfsubject={\papertitle},
  pdfkeywords={\paperkeywords},
  pdffitwindow=true, 
  breaklinks=true, 
  colorlinks=false, 
  pdfborder={0 0 0} 
}

\begin{document}

\title{\papertitle\footnote{Preprint of: \papercitation}}
\author[]{\myname}
\affil[]{\myaffiliation}
\date{\paperdate}

\maketitle

\begin{abstract}

  Urban park accessibility shapes mental health, physical well-being, and community cohesion. Measuring it is thus central to urban and landscape planning. Although researchers have debated the impacts of different accessibility measures on analysis results, scant attention has been paid to variability from data source selection. Do the results of accessibility analyses vary depending on the dataset used to define park locations and attributes? We compare four widely used geospatial data sources---OpenStreetMap, Esri, SafeGraph, and The Trust for Public Land---by validating their data and measuring park access across two cities. These datasets purport to cover the same spatial features, but our findings reveal wide discrepancies in accessibility results, driven by differences in dataset completeness, coverage, and accuracy between the sources. We provide guidelines for selecting and carefully validating data for accessibility analyses, as these decisions directly impact the accuracy and equity of downstream urban park planning and decision-making.

\end{abstract}
\vfill
\section{Introduction}

Cities exist because people seek access to things they need or want. Measuring this accessibility helps us understand urban patterns, processes, and outcomes. For example, access to urban parks and green spaces has links to mental health, physical well-being, and community cohesion \citep{cohen_contribution_2007, wood_public_2017}. Across the urban planning, transport geography, and public health literatures, researchers have studied park accessibility empirically as well as methodologically to improve the conceptualization and measurement of urban park access for more equitable urbanization and evidence-informed planning \citep{sherer_benefits_2006, wolch_urban_2014, world_health_organization_urban_2017}.

However, despite the rich empirical literature around park accessibility, there is little consensus on how to measure it \citep{wolch_urban_2014}. Many methodological papers address indicators and analytics from the perspectives of theory or calculation, but few consider the challenges of the underlying source data themselves. This gap creates two problems. First, accessibility analyses depend on well-defined origins, destinations, and impedances between them, but these are defined inconsistently from one study to another. Second, the geospatial data used to measure these values vary across sources in terms of how parks are defined, represented, and described. While some data sources provide detailed, comprehensive representations of urban parks, others omit key features, misclassify spaces, or fail to capture informal public green spaces. Further, the source data themselves rarely map 1:1 to the precise theoretical construct of a \enquote{park} (which is itself a fuzzy real-world object) on which a specific study relies. Accordingly, this study asks: do the results of park accessibility analyses vary depending on which park data source you choose?

We compare four widely used geospatial datasets on park locations and attributes---from OpenStreetMap (OSM), Esri, SafeGraph, and The Trust for Public Land (TPL)---across two different cities. We manually validate the source data, then model access to parks using each source and compare the results. Our findings reveal substantial discrepancies in measured accessibility, driven by differences in data source completeness and coverage. From this, we develop guidelines for planners selecting and validating data for accessibility analyses. No dataset is perfect. Analysts must account for each's limitations when measuring urban parks, triangulate across data sources, and ground-truth their data for more accurate urban accessibility analyses.

\section{Background}

Urban residents value access to parks for both recreation and health \citep{sherer_benefits_2006,wolch_urban_2014,world_health_organization_urban_2017}. Parks offer a variety of amenities, but in this study we focus on parks that provide green spaces for physical activities that improve long-term health outcomes and promote mental well-being \citep{cohen_contribution_2007, wood_public_2017}. In the literature, such parks are typically required to be >2 acres, ensuring that their spatial characteristics can adequately support meaningful physical activity and community use \citep{annerstedt_van_den_bosch_development_2016, boeing_using_2022,turner2025internationally}.

Planners pursuing the benefits of such parks require unbiased methods to measure public accessibility to these green spaces to understand the status quo and design spatial interventions. In the context of public health, measuring access to green spaces often involves assessing impedance: the cost or difficulty of reaching a destination, incorporating factors like distance, time, monetary cost, cognitive burden, or some combination of these \citep{mowen2005change,kaczynski_association_2008,orstad_park_2020,wang_access_2021,gong_modeling_2023,el-murr_walking_2021,schipperijn_access_2017,maas_green_2006,browning_within_2017, mitchell_effect_2008, helbich_more_2018, wu_impact_2024}. Green space itself is construed in various ways: as visually perceived greenery, behaviorally experienced environments, or physical publicly accessible green space \citep{james_exposure_2016,cohen-cline_access_2015,oregan_biophilic_2021,larkin_evaluating_2019,mears_mapping_2021,pirotti_analysis_2021,zabelskyte_patterns_2022,schwartz_visitors_2019}. This study focuses on the latter---measuring physical access to green space represented by public parks and recreation areas.

\subsection{Common Park Accessibility Measures}

The distance to the nearest park is a fundamental measure of accessibility, as closer proximity is associated with higher park usage \citep{toftager_distance_2011}. To account for practical accessibility, many measures consider network-constrained travel---calculating walking distance or time along actual paths---to reflect real-world mobility \citep{wang_access_2021,orstad_park_2020,gong_modeling_2023,boeing_travel_2026,kaczynski_association_2008}. For instance, the number of parks accessible within a given walking distance or time helps capture potential exposure and usage \citep{el-murr_walking_2021, schipperijn_access_2017}. Similarly, the accessible park area measures the total green space within a buffer, where larger areas correlate with greater usage \citep{maas_green_2006,browning_within_2017}. Supply-demand measures assess green space availability per capita or neighborhood unit, with the aim of equitable distribution \citep{wu_impact_2024}, as sufficient green space per person supports fair access and health equity \citep{mitchell_effect_2008,helbich_more_2018}.

\subsection{Common Park Accessibility Data}

Accessibility analyses depend on well-defined origins and destinations. Origins represent the points from which people travel to destinations, often defined as residences \citep{maas_green_2006,toftager_distance_2011,gong_modeling_2023}, tract centroids or boundaries \citep{browning_pad-us-ar_2022}, or nearby street intersections \citep{jang_urban_2020}.

Representing the destination amenities---in this case parks---is more variable. OpenStreetMap (OSM), a popular crowd-sourced mapping platform, provides a vast collection of geographic data, including parks and green spaces such as playgrounds, grassy road medians, and nature reserves, some of which may lack public accessibility/utility. In contrast, proprietary datasets, such as those provided by Esri and SafeGraph, compile data through more formal channels. Esri's US Parks dataset, for example, aggregates proprietary sources including the Overture Maps Foundation, OSM, partner datasets, and sensor-derived observations to capture natural or semi-natural areas designated for public recreation and conservation. Other datasets, such as those from the Trust for Public Land (TPL), combine multiple data sources, including local governments and crowd-sourced updates from tools like ParkReviewer. Despite the intended comprehensive nature of such datasets, their accuracy and completeness could vary depending on the location and data collection methods employed.

Parks vary widely in size, shape, and spatial complexity, which directly influences how people access and use them. Datasets represent parks either as points or polygons, and this choice significantly affects accessibility analysis. While access to a small pocket park could be reasonably approximated with a point feature, access to large urban parks such as Central Park in New York or Griffith Park in Los Angeles would be poorly represented. Point data, representing parks as single coordinates, fail to capture their spatial extent, often omitting critical details such as actual entrances or internal layouts \citep{psyllidis_points_2022}. This can lead to unrealistic access measurements, especially when access is calculated from centroids, abstract simplifications that rarely reflect how users actually enter and navigate parks. In contrast, polygons provide spatial footprints, delineating shape, area, and boundaries, which better represent real-world interaction. However, even with polygons, access is often measured from centroid or edge points, overlooking the actual connection points to the street network \citep{macfarlane_beyond_2022}.

OSM, Esri, TPL, and SafeGraph represent parks as polygons. However, even these data sets may lack critical details, such as entrance locations and opening hours. The most accurate park data would ideally come from administrative sources. However, the quality of these datasets can vary significantly depending on the size of the city, as larger or wealthier urban areas may be better resourced. Furthermore, even in well-documented regions, datasets are not always up to date, as new parks, renovations, or changes in accessibility may not be promptly reflected.

\subsection{Research Problem}

The literature has established the importance of urban green space access for public health and well-being, but measuring such access remains problematic: researchers and practitioners often do not understand whether systematic biases exist among different data sources that represent these spaces \citep{wolch_urban_2014}. Such biases arise from scope, quality, and mismatches between theoretical definitions of spatial features and their representations in the data. This problem has only grown in recent years as a plethora of parks data sources has emerged. These data sources rarely advertise themselves as biased or incomplete, so a naïve baseline assumption would be that they are thus interchangeable with each other. In other words, scholars and practitioners simply do not know \textit{a priori} how their accessibility analyses depend on their input data source. Worse, many studies rely on specific data sources without justifying their selection.

\section{Methods}

\begin{figure}
  \centering
  \includegraphics[width=1\linewidth]
  {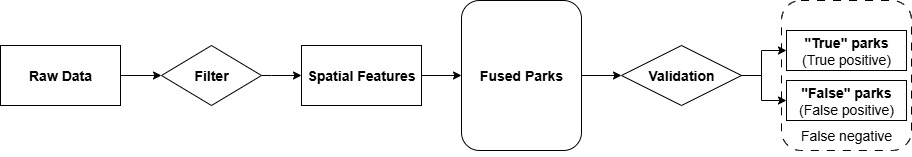}
  \caption{Workflow overview: we collect, filter, fuse, and validate the data.}\label{fig:workflow}
\end{figure}

To address this problem, this study asks: do the results of park accessibility analysis vary depending on the data sources used? We hypothesize that different spatial data sources---all purporting to provide equivalent spatial features, yet potentially varying in undocumented ways in coverage, accuracy, and digitization standards---result in different measurements of accessibility to parks in a city. To test this, we measure access to parks using several data sources. We manually validate the source data then compare the accessibility analyses' results to understand how they vary based on the input data. \autoref{fig:workflow} illustrates this entire process from data collection to validation.

\subsection{Data}

\subsubsection{Defining Parks}

The first challenge in conducting a park accessibility analysis is defining which spatial features to include as destination amenities. Here, we follow a common standard from the public health literature to define \enquote{parks} as large public urban green spaces: we define \textit{large} as a minimum of 2 acres so that the spatial features can support meaningful physical activity \citep{annerstedt_van_den_bosch_development_2016,boeing_using_2022,turner2025internationally}. We define \textit{public} as publicly accessible at least some of the time, meaning we include spatial features like school fields that may have restricted hours. We define \textit{urban} as being located within a core city's municipal boundaries. We define \textit{green} by excluding fully-paved spatial features such as basketball courts. Finally, we define \textit{space} as areas, thus excluding zero- or one-dimensional spatial features. Henceforth, we refer to these carefully defined and validated amenities as \enquote{true} parks and the initial unvalidated spatial features in our datasets as \enquote{spatial features.}

\subsubsection{Data Source Selection}

To capture some variation in city size, region, socioeconomics, and demographics, we analyze two US cities---Portland, Oregon and Buffalo, New York---as validation study sites. Each of these two study sites comprises the area within the municipal boundary buffered by 2.4 km to mitigate perimeter effects and ensure that we capture spatial features within a walkable distance of residents near the city limits \citep{colclough_mapping_2010}. Spatial features that extended beyond this walkable distance boundary were clipped at the buffer edge to maintain consistency in the study area. Buffalo's buffer extends into Canada, but we exclude Canadian spatial features due to border crossing constraints.

The next challenge is identifying which data source to use. Because our study areas span multiple municipalities intersecting a 2.4 km buffer around each core city, we exclude inconsistently available individual municipality data from our analysis, and instead compare four of the most commonly used spatial data sources: OSM, TPL, Esri, and SafeGraph \citep{williams_parks_2020}. These data sources provide polygon features, which are necessary to apply our minimum size threshold and to measure park boundaries. Thus we must exclude other common data sources like Foursquare, Google, Here, TomTom, Yelp, and Mapbox as they provide park features either as points or an inconsistent mix of points and some polygons.

\subsubsection{Data Collection}

The next challenge is identifying which spatial features to use from these sources. In cases where parks are explicitly described \textit{and} match our own definition, we use the data directly. Otherwise, we must carefully filter parks based on attribute data to align with our precise definition for this study.

We use the TPL dataset from the ParkServe database, which includes \enquote{a wide variety of parks, open-access schoolyards, trails, and natural areas} that are publicly accessible spaces \citep{Trust_for_Public_Land_web}. They explicitly exclude private or restricted spaces like parks in gated communities, private golf courses, and school playgrounds without formal agreements to ensure their database consists of areas open to the public. To precisely match our definition of parks used in this study, we filter this dataset by park areas at least 2 acres, and within our municipal study sites. Our remaining criteria are already met by these data: they are publicly accessible green areas. Next, we download the Esri dataset from ArcGIS Online \citep{esri_us_nodate}. To precisely match our definition of parks used in this study, we filter this dataset by park areas at least 2 acres, and within our municipal study sites. We can perform no additional filters on this dataset because they provide no attribute data: thus this represents the best possible good-faith effort for an apples-to-apples comparison aligning with our definition of parks.

Next, we use the OSMnx package \citep{boeing_modeling_2025} to collect polygonal spatial features from OSM tagged with these specific key-value tags to align with our definition:

\begin{enumerate}

  \item landuse: \enquote{common}, \enquote{conservation}, \enquote{forest}, \enquote{garden}, \enquote{leisure}, \enquote{park}, \enquote{recreation\_ground}, \enquote{sport}, \enquote{trees}, \enquote{village\_green}, \enquote{winter\_sports}, \enquote{wood}, \enquote{dog\_park}, \enquote{nature\_reserve}, \enquote{off\_leash}, \enquote{sports\_centre}, \enquote{riverbank}, \enquote{beach}

  \item boundary: \enquote{national\_park}, \enquote{nature\_reserve}, \enquote{forest}, \enquote{state\_forest}, \enquote{state\_park}, \enquote{regional\_park}, \enquote{park}, \enquote{county\_park}, \enquote{protected\_area}

  \item leisure: \enquote{dog\_park}, \enquote{park}, \enquote{nature\_reserve}

\end{enumerate}

Finally, we collect the SafeGraph data by downloading the dataset and selecting only spatial features classified as \enquote{Nature Parks and Other Similar Institutions} in the SUB\_CATEGORY field.

\subsection{Validation}

We assign a unique identifier to each spatial feature from each data source. We then identify candidate matches across the four data sources based on spatial proximity and name similarity, producing a preliminary \enquote{fused parks} dataset in which each row represents a potential park that may correspond to multiple features across datasets. To operationalize name similarity in a reproducible manner, we compute pairwise fuzzy string similarity scores (derived from Levenshtein distance across park names) and retain candidate matches only when all pairwise comparisons exceed a threshold of 70 (i.e., roughly 70\% similar after normalization). We choose this threshold after iterative testing as a conservative criterion to capture only clearly matching cases at this initial stage.

To conduct the validation, two independent validators inspect each of these candidate spatial features using Google Maps satellite imagery, Google Street View, Google Reviews, official websites, and other publicly available information. Importantly, the fuzzy matching procedure serves only to reduce the burden of manual inspection and does not determine the final park entities. Validators make the final decision on whether features represent the same park, including in cases where names are missing, inconsistent, or only partially similar (e.g., borderline similarity scores). Each validator evaluates whether the spatial feature is a park (true or false), and we record specific issues, such as cases where a park incorrectly merges multiple distinct areas under a single name, along with additional notes. To identify potential false negatives (i.e., parks that exist but are not present in any dataset), we conduct an additional manual visual validation on a random sample 6 tracts in Portland and 4 in Buffalo. The validator uses satellite imagery and unfiltered spatial features from each data source to manually inspect each tract and identify any potential parks missing in the initial spatial feature datasets.

\subsubsection{Agreement Between Raters}

We calculate Cohen's kappa coefficient to measure inter-rater reliability and determine the strength of agreement between our two validators, where $\kappa$ = 1 means complete agreement between validators and $\kappa$ = 0 means no agreement between raters. In the social sciences, $\kappa \geq$  0.6 is a standard threshold for \enquote{substantial agreement} \citep{landis_application_1977}.

\subsubsection{Agreement Between Data Sources}

We calculate the precision, recall, and F1 score for each data source in each city to analyze how they perform relative to each other. Precision is the true positive count divided by the sum of true positives and false positives. This ranges from 0 to 1, with 1 indicating that all the spatial features in a data set were identified by the validators as \enquote{True} parks.

Recall is the true positive count divided by the sum of true positives and false negatives. Because parks missing from all sources cannot be detected, a false negative for a given source is defined here as a park absent from that source but present in at least one other source and validated by validators as a \enquote{True} park. Recall measures the proportion of \enquote{True} parks that each dataset successfully captured as \enquote{True.} This ranges from 0 to 1, with 1 indicating that the dataset captured every \enquote{True} park. Finally, the F1 score is the harmonic mean of precision and recall, providing an overall measure of accuracy.

\subsection{Accessibility Analysis}

We use the set of all street intersections/dead-ends (i.e., nodes) on OSM as origins for the accessibility analysis. Using 2020 census and LEHD block-level data, we find that node counts are consistently and positively correlated with both population and employment within tracts, making them a useful proxy for human presence. Unlike residence-based weighting, intersections provide a more balanced spatial representation of where people spend time, including areas with high employment but low residential population. For each such node, we pre-compute 15-minute walking isochrones assuming a speed of 4.8 km/hour, adopting the \enquote{15-minute cities} paradigm of a convenient distance at an average walking speed \citep{evans_accessibility_2009, tennoy_walking_2022, colclough_mapping_2010}. We calculate walkable network-based access for its emphasis on practical accessibility rather than mere spatial proximity \citep{el-murr_walking_2021}, using OSM data on streets and OSMnx for modeling. Next, for each isochrone, we calculate three key accessibility indicators separately for each of the four data sources using the original, unvalidated features. For the validated composite dataset, we include only spatial features that both validators agreed were parks, excluding all cases of disagreement.

First, \textit{Accessible Park Count} calculates, for each isochrone, the count of park entities accessible within a 15-minute walking distance, providing a simple count-based measure of accessibility. Second, \textit{Accessible Park Area} calculates the area (in km\textsuperscript{2}), of all intersecting park polygons within a 15-minute walking distance from that origin, offering an area-based measure of potential green space access. Third, \textit{Distance to Nearest Park} calculates the distance to the nearest park from that origin, along real-world streets and footpaths, offering a distance-based measure of nearest green space, measured in km.

With these per-node isochrone-based accessibility calculations, we then aggregate each of these three indicators to more interpretable geographical units. We first aggregate to the census tract level by calculating the mean value of all origin nodes within each tract. To compare park accessibility between data sources, we measure and visualize these tract-level indicator means across each data source for both cities. To make city-level comparison between data sources, we use the median of tract-level mean values for Accessible Park Count and Accessible Park Area, and the mean of tract-level mean values for Distance to Nearest Park.

Finally, we explain the planning implications of data source selection in terms of identifying \enquote{underserved} census tracts. We define a tract as underserved if the distance to its nearest park exceeds 300 meters, aligning with \citet{world2016urban} standards. For each data source, we compute two proportions: (1) the proportion of underserved tracts; (2) the proportion of tracts that shift from underserved to not (or vice versa) compared to the validated composite dataset, representing the misclassification rate. Ideally, each dataset would have a similar proportion of underserved tracts and a near-zero misclassification rate when compared to the validated composite dataset.

\section{Findings}

\subsection{Validation}

\subsubsection{Agreement Between Data Sources}

\begin{figure}
  \centering
  \includegraphics[width=1\linewidth]{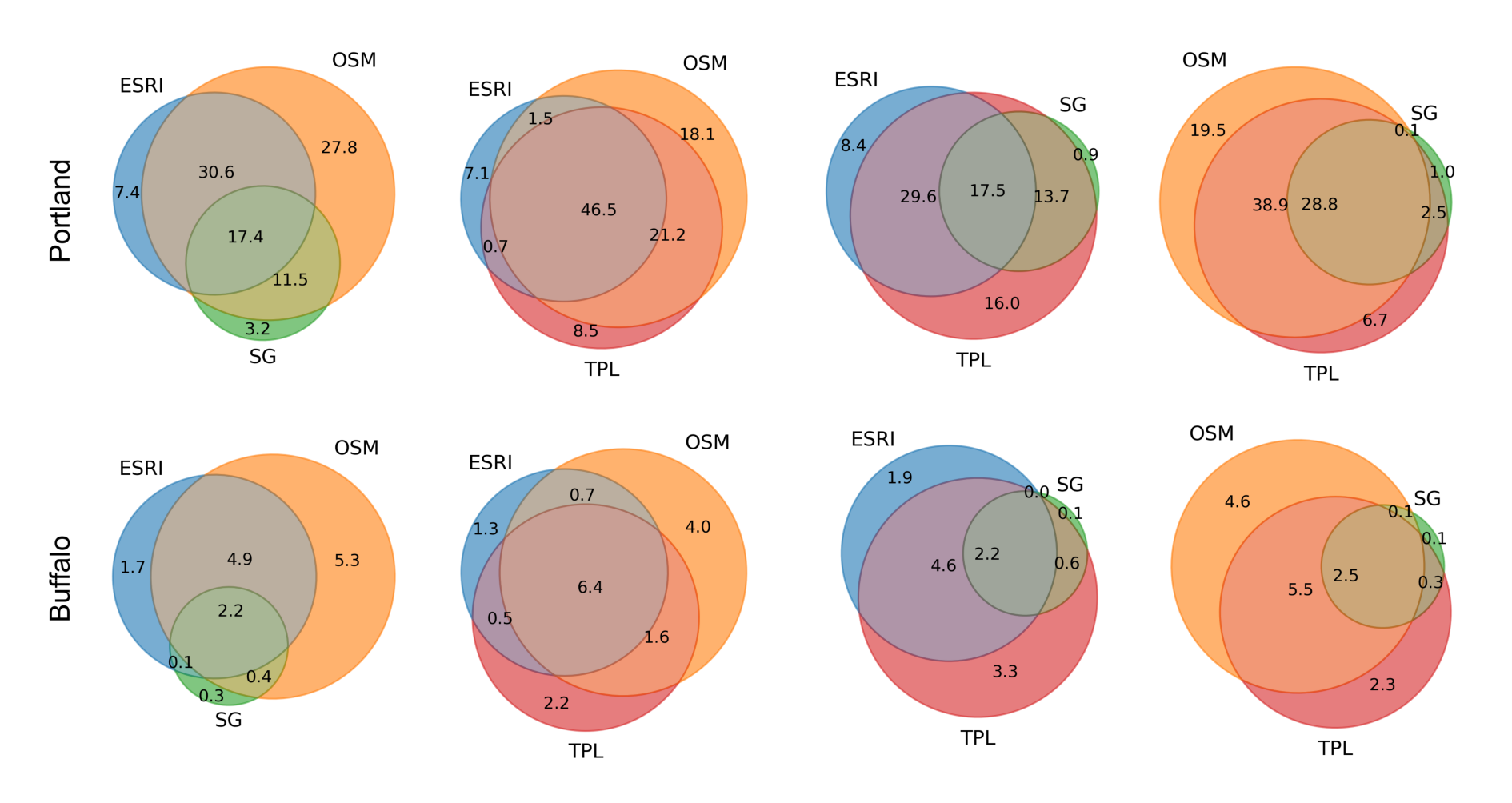}
  \caption{Venn diagrams showing overlap (km\textsuperscript{2}) of spatial feature areas among Esri, OSM, TPL, and SafeGraph. Values reflect shared and unique areas across three sources per city.}\label{fig:venn}
\end{figure}

The manual visual validation identifies no false negatives missing in the initial data collection. Figure~\ref{fig:venn} shows how spatial feature area overlaps in each city between the different datasets. Because of the nature of Venn diagrams, it is infeasible to display four datasets in a single two-dimensional figure, so they are shown in groups of three datasets at a time. For example, in Portland, Esri and OSM share 48 km\textsuperscript{2}, Esri and SafeGraph share 17.4 km\textsuperscript{2}, and OSM and SafeGraph share 28.9 km\textsuperscript{2} of spatial feature areas. Only 17.4 km\textsuperscript{2} of spatial feature areas are shared by all three---Esri, OSM, and SafeGraph---datasets. Esri has 7.4 km\textsuperscript{2}, OSM has 27.8 km\textsuperscript{2}, and SafeGraph has 3.2 km\textsuperscript{2} of spatial feature area that is not shared with the other two datasets.

Across both cities, before validation, we find that the OSM dataset contains the most park area (100.02 km\textsuperscript{2}), followed by TPL (87.55 km\textsuperscript{2}), Esri (64.52 km\textsuperscript{2}), and SafeGraph (35.27 km\textsuperscript{2}). TPL and OSM datasets have the most overlap in both cities (67.7 km\textsuperscript{2} in Portland and 7.9 km\textsuperscript{2} in Buffalo), while Esri and SafeGraph have the least overlap (17.4 km\textsuperscript{2} in Portland and 2.1 km\textsuperscript{2} in Buffalo). The OSM data overlap the most with the other data sources, while SafeGraph's limited data overlap the least.

\subsubsection{Agreement Between Raters}

We find that there is substantial agreement in validation results between our 2 raters across both Portland and Buffalo. Raters agree on 1627 out of 1710 (95.2\%) total spatial features. Both raters record \enquote{True} for 1378 (80.6\%), and both record \enquote{False} for 249 (14.6\%). The raters disagree on 83 spatial features (4.8\%). For 49 spatial features (2.9\%), Rater 1 records \enquote{True}, and Rater 2 records \enquote{False}. For 34 spatial features (about 2.0\%), Rater 1 records \enquote{False} and Rater 2 records \enquote{True}. Cohen's $\kappa$ is 0.868 in Portland ($n=1264$) and 0.735 in Buffalo ($n=446$): both coefficients exceed the $\kappa=0.6$ threshold for substantial agreement and have $p<0.001$.

\subsubsection{Validation Results Across Data Sources}

Across both cities, raters validate 451 out of 620 spatial features and 71.59 km\textsuperscript{2} out of 100.02 km\textsuperscript{2} of spatial feature area as \enquote{True} in OSM's dataset; 410 out of 503 spatial features and 67.78 km\textsuperscript{2} out of 87.55 km\textsuperscript{2}  spatial feature area as \enquote{True} in TPL's dataset; 328 out of 359 spatial features and 53 km\textsuperscript{2} out of 64.52 km\textsuperscript{2} of spatial feature area as \enquote{True} in Esri's dataset; and 189 out of 235 spatial features and 22.69 km\textsuperscript{2} out of 35.27 km\textsuperscript{2} of spatial feature area as \enquote{True} in SafeGraph's dataset.

Analyzing Precision, Recall, and F1 Scores reveals significant disagreement among the four data sources, with Esri, OSM, and TPL generally performing better overall than SafeGraph. In Portland, Esri has the highest precision score (0.949), meaning that the largest share of the features it identified were confirmed as true parks, followed by TPL (0.836), SafeGraph (0.833), and OSM (0.799), while OSM has the highest recall score (0.900), identifying the largest proportion of parks compared to the total number of \enquote{True} parks, followed by TPL (0.820), Esri (0.621), and SafeGraph (0.388). In Portland, OSM has the highest F1 score (0.847), followed by TPL (0.828), Esri (0.751), and SafeGraph (0.530).

In Buffalo, Esri again has the highest precision score (0.882), followed by TPL (0.847), SafeGraph (0.815), and OSM (0.678), while OSM has the highest recall score (0.836), identifying the largest proportion of parks compared to the total number of existing parks, followed by TPL (0.782), Esri (0.740), and SafeGraph (0.333). In Buffalo, TPL has the highest F1 score (0.813), followed by Esri (0.805), OSM (0.749), and SafeGraph (0.473).

Across both cities, some data sources consistently perform well, with Esri's high precision scores, OSM's high recall scores, and TPL's strong overall performance. In contrast, SafeGraph has the lowest recall and F1 scores in both cities, indicating that it misses many parks.

\subsection{Accessibility Analysis Findings}

Table~\ref{tab:table_1} summarizes the accessibility indicator values across data sources and cities. OSM shows relatively high accessibility across both cities, with high park counts and park areas identified as accessible within 15-minute walking distance. By the same indicators, Safegraph consistently performs poorly, with the least park accessibility in both cities.

\begin{table}[h!]
  \centering
  \begin{tabular*}{\textwidth}{@{\extracolsep{\fill}}lcccccc@{}}
    \toprule
    \multirow{2}{*}{~} &
    \multicolumn{2}{c}{\thead{Accessible\\Park Count}} &
    \multicolumn{2}{c}{\thead{Accessible\\Park Area}} &
    \multicolumn{2}{c}{\thead{Distance to\\Nearest Park}} \\
    \cline{2-7}
    \addlinespace[3pt]
    & Portland & Buffalo
    & Portland & Buffalo
    & Portland & Buffalo \\
    \midrule
    OSM        & 3.30 & 3.27 & 0.089 & 0.091 & 0.38 & 0.41 \\
    TPL        & 3.23 & 2.62 & 0.091 & 0.092 & 0.38 & 0.45 \\
    Esri       & 2.46 & 3.03 & 0.083 & 0.085 & 0.48 & 0.44 \\
    SafeGraph  & 1.55 & 1.30 & 0.035 & 0.022 & 0.70 & 0.82 \\
    \bottomrule
  \end{tabular*}
  \caption{Comparison of park accessibility indicator values by unvalidated data source and city. Indicators and units are described in the text.}\label{tab:table_1}
\end{table}

\subsubsection{Portland}

\begin{figure}
  \centering
  \includegraphics[width=\linewidth]{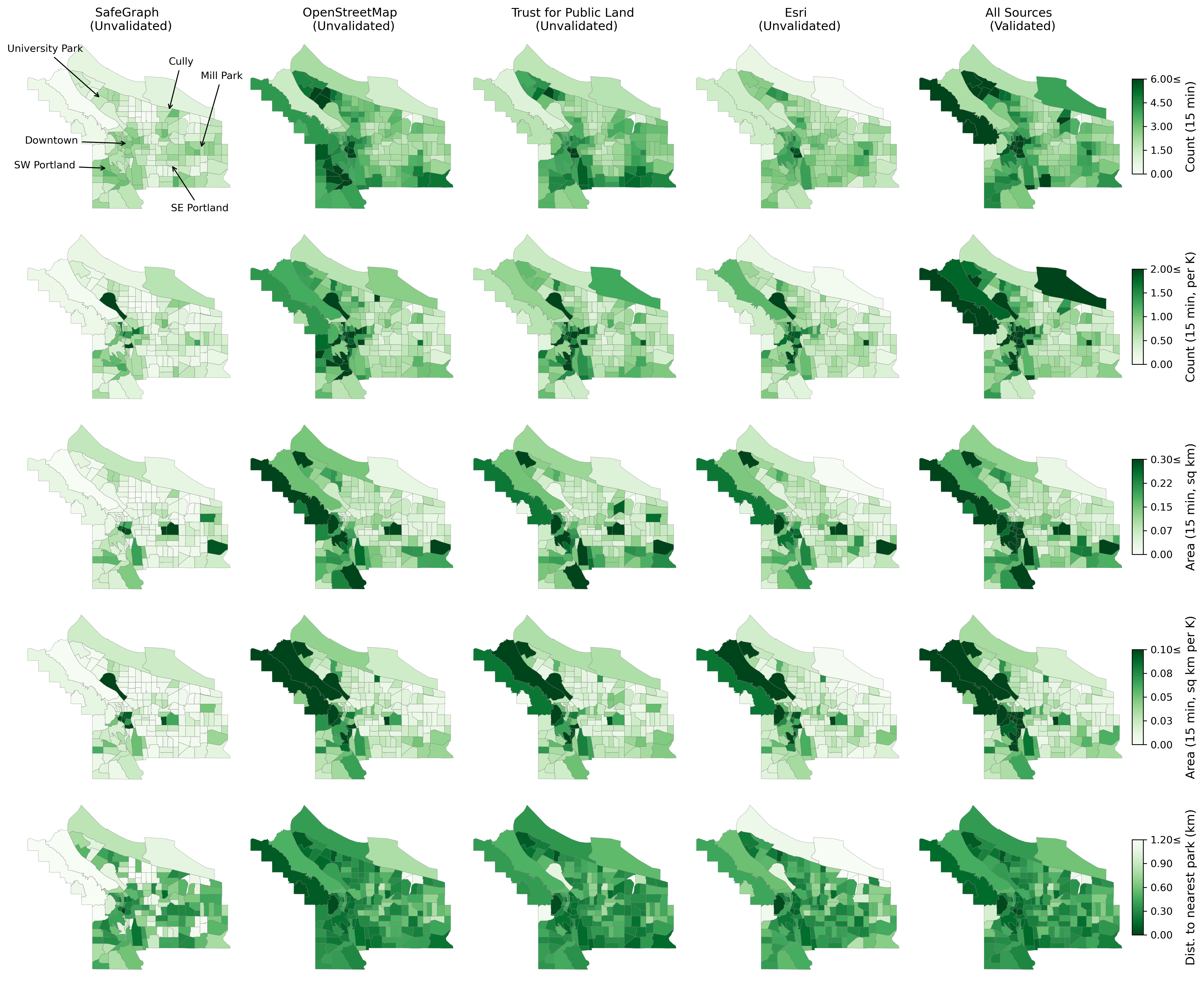}
  \caption{Census tract-level park accessibility in Portland across data sources. Comparison of park accessibility indicators---including park count and area within 15 minutes of travel, and distance to the nearest park---based on four unvalidated datasets (SafeGraph, OSM, TPL, Esri) and a validated composite dataset.}\label{fig:portland}
\end{figure}

As illustrated by \autoref{fig:portland}, accessibility measures vary depending on both the indicator and dataset. First we look at Portland. By the Accessible Park Count indicator, OSM identifies 3.30 parks as accessible within 15 minutes walking. Within the same distance, TPL identifies 3.23 parks, Esri identifies 2.46 parks, and SafeGraph identifies 1.55 parks. By the Accessible Park Area indicator, OSM identifies 0.089 km\textsuperscript{2} of park area as accessible within 15 minutes walking. Within the same distance, TPL identifies 0.091 km\textsuperscript{2}, Esri data identifies 0.083 km\textsuperscript{2}, and SafeGraph identifies 0.035 km\textsuperscript{2}. By the Distance to Nearest Park indicator, TPL and OSM data indicate that the average distance to the nearest park is 0.38 km, while Esri data indicates 0.48 km, and Safegraph data indicates 0.70 km.

The population-normalized measures show a similar pattern. Per 1,000 residents, the validated composite dataset provides access to 0.82 parks and 0.0230 km\textsuperscript{2} of park area. Among the unvalidated sources, OSM reports 0.81 parks and 0.026 km\textsuperscript{2} per 1,000 residents, followed by TPL with 0.78 parks and 0.025 km\textsuperscript{2}, Esri with 0.63 parks and 0.022 km\textsuperscript{2}, and SafeGraph with 0.37 parks and 0.009 km\textsuperscript{2}.

SafeGraph identifies 87.8\% of tracts as underserved whereas only 57.9\% are so in the validated composite dataset. 34.8\% of tracts are misclassified (i.e., underserved classified as non-underserved, or vice versa) in the SafeGraph data when compared to the validated composite dataset. OSM (60.4\%), TPL (64.6\%), and Esri (72.0\%) show closer alignment with the proportion of underserved tracts in the validated composite dataset, and 12--15\% of tracts are misclassified in those three datasets.

\subsubsection{Buffalo}

\begin{figure}
  \centering
  \includegraphics[width=0.95\linewidth]{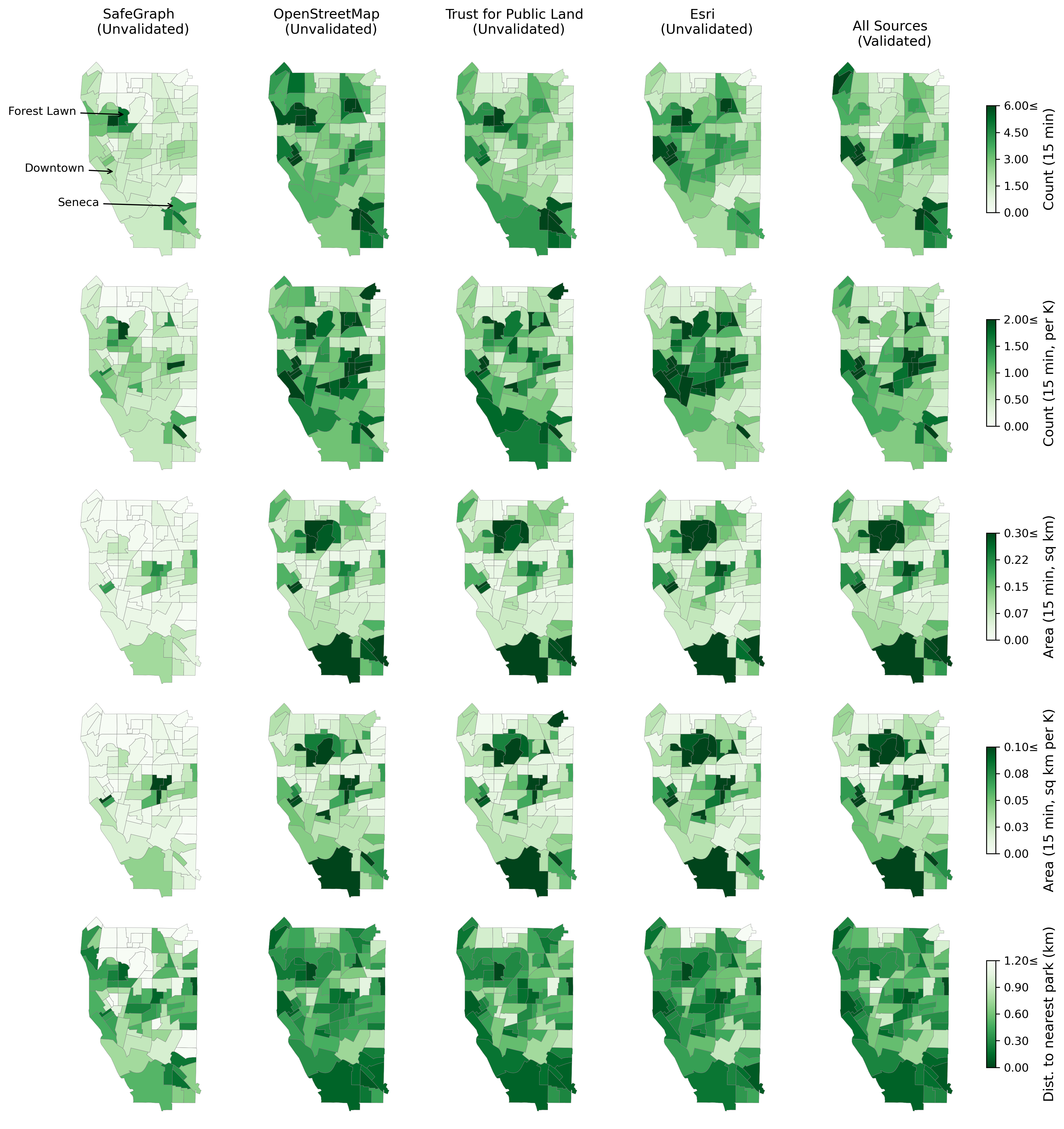}
  \caption{Census tract-level park accessibility in Buffalo across data sources. Comparison of park accessibility indicators---including park count and area within 15 minutes of travel, and distance to the nearest park---based on four unvalidated datasets (SafeGraph, OSM, TPL, Esri) and a validated composite dataset.}\label{fig:buffalo}
\end{figure}

Buffalo's accessibility measures, shown in \autoref{fig:buffalo}, also show inconsistencies between indicators and datasets. By the Accessible Park Counts indicator, OSM identifies 3.27 parks as accessible within 15 minutes walking. Within the same distance, TPL identifies 2.62 parks, Esri data identifies 3.03 parks, and SafeGraph identifies 1.30 parks. By the Accessible Park Area indicator, OSM identifies 0.091 km\textsuperscript{2} of park area as accessible within 15 minutes walking. Within the same distance, TPL identifies 0.092 km\textsuperscript{2}, Esri data identifies 0.085 km\textsuperscript{2}, and SafeGraph identifies 0.022 km\textsuperscript{2}. By the Distance to Nearest Park indicator, OSM data indicates that the average distance to the nearest park is 0.41 km, while TPL data indicates 0.45 km, Esri data indicates 0.44 km, and Safegraph data indicates 0.82 km.

The second and fourth rows of \autoref{fig:buffalo} present the population-normalized measures. Per 1,000 residents, the validated composite dataset provides access to 0.97 parks and 0.034 km\textsuperscript{2} of park area. Among the unvalidated sources, OSM reports 1.18 parks and 0.033 km\textsuperscript{2} per 1,000 residents, followed by TPL with 0.99 parks and 0.029 km\textsuperscript{2}, Esri with 0.96 parks and 0.032 km\textsuperscript{2}, and SafeGraph with 0.47 parks and 0.009 km\textsuperscript{2}.

SafeGraph identifies 87.9\% of tracts as underserved whereas only 62.6\% of tracts are so in the validated composite dataset. 31.9\% of tracts are misclassified (i.e., underserved classified as non-underserved, or vice versa) in the SafeGraph data when compared to the validated composite dataset. Again OSM, TPL, and Esri show closer alignment with the proportion of underserved tracts in the validated composite dataset (all ranging 64.8--67.0\%), and 11--15\% of tracts are misclassified in those three datasets.

\section{Discussion}

When researchers and practitioners seek to analyze urban accessibility, they must identify data to represent origins, destinations, and impedances. Origins are often represented as residences or abstracted to the tract-level or network nodes. Impedances are often modeled from the street network, via distance, travel time, or similar. But destinations require a dataset of amenity locations. In this study, we examine data from OSM, TPL, Esri, and SafeGraph---all of which commonly appear in the literature. Each purports to provide a dataset representing the same real-world entities: \enquote{parks,} so our hypothetical analyst should be able to calculate the same accessibility measures from each.

To test this, we asked: how do different geospatial data sources influence the measurement of urban park accessibility? Our findings support the hypothesis that common accessibility measurements---such as park count, area, and distance to nearest---vary depending on the data source used. These variations are not trivial: they fundamentally change our understanding of where parks are located and who has access to them. A planning practitioner reading the descriptions of any of these data sources would presumably feel confident that they intend to represent interchangeable sets of parks.

This could lead us astray for several reasons. First, the definition of \enquote{park} varies not only according to the researcher’s theoretical framework, but also across data sources' digitization standards. Not all data sources clearly define what constitutes a park. For example, some data sources include roadway medians or grassy lots, while others exclude natural areas or school fields. Thus it is critically important that they provide relevant attribute data for filtering to align with the theoretical construct under study and to operationalize study-specific definitions. For example, while Esri's dataset employs a broad definition of parks, it lacks the necessary attributes to support detailed filtering.

Furthermore, parks are inherently \textit{areas} of interest, not \textit{points} of interest, making polygonal/areal analysis essential to understand the scale and extent of the amenity. Yet many datasets represent them as representative points instead of areas, complicating accessibility analyses. Some data sources disaggregate individual adjacent parks, whereas others aggregate them, biasing counts of discrete park entities. Some large nature parks, such as Forest Park in Portland, comprise many small component parks: there is no objectively right way to represent these. OSM represents these component parks as separate spatial features, whereas Esri and TPL represent them all as one mega-park. In addition, TPL data do not use multipolygon geometry types for individual park entities spanning multiple disjoint areas. TPL also includes some golf courses in Portland, though their documentation explicitly states that they do not. SafeGraph's park polygons often misrepresent a park or athletic field's boundary as a single building within the confines, rather than the park/field itself. For some parks, TPL only includes the trail path through it, whereas other data sources include the surrounding public green space.

Beyond the data sources themselves, validation is challenging---particularly for less-formal green spaces such as nature parks, natural areas, and reserves---when assessing public access. Real-world ambiguity can complicate analyses since parks are fuzzy objects. For example, sometimes only a small portion of a natural area is usable by the public, but the whole area is identified as a park: there is no objective right or wrong spatial definition here, rather just the subjective alignment with a specific study's representational needs. Some parks limit public access to scheduled open hours, such as fields open to the public after school and on weekends, or small hospital parks publicly accessible during visiting hours. Information on these temporal constraints is not always readily available, presenting further challenges.

In this study, our validation team used Google reviews, Google Street View imagery, and official park websites to make informed judgments regarding whether the spatial feature allowed public access or not. This strength of our study is also an important limitation: this is time- and labor-intensive, requiring substantial resourcing for careful analytics, which may pose a challenge for resource-constrained practitioners. Fortunately, this validation workflow could potentially be automated with AI in coming years (though AI itself could introduce a new source of bias that requires careful human review). In the meantime, this represents a trade-off between measurement precision and analytical effort. Manual validation is labor-intensive but reveals biases that would otherwise remain hidden. We argue that planners and practitioners should not take spatial datasets at face value, but should cross-check multiple sources and verify that the data actually represent the features relevant to their analysis.

While each of these data sources represent parks, small differences in methods, digitization, and representation complicate downstream interpretations and analyses. Our results show how datasets do not provide purely neutral interchangeable facts---they come with inherent biases that shape urban planning knowledge in subtle but important ways. Practitioners must recognize that even for-profit datasets marketed as comprehensive, such as expensive SafeGraph data (which performed particularly poorly in our validation), have limitations. The idealized notion of a \enquote{perfect} dataset does not exist. Rather, some are better for certain analyses than others.  For example, in Portland, TPL and OSM datasets offer more complete coverage of natural areas and nature preserves than SafeGraph or Esri, making them a better choice for studies focusing on these types of green spaces. The key is aligning a study's theoretical needs with the best representation in data. Dataset selection also impacts which communities are deemed underserved. In particular, SafeGraph's incomplete coverage suggests larger access deficits, a data bias that risks misallocating resources away from genuinely underserved areas.

Beyond park accessibility itself, these findings have broader implications for empirical research that relies on mobility data---particularly studies using SafeGraph to analyze park visit patterns \citep[e.g.,][]{jay2022use, sun2024seasonal, zhao2023change}. While SafeGraph has become a popular data source for understanding human movement, our validation demonstrates that its park geometries are often incomplete or misrepresented, leading to potential biases in measures of visitation frequency, user demographics, or temporal activity patterns. Such distortions can bias downstream policy or equity analyses that depend on accurate representations of spatial features.

We call attention to the potential of these differences between data sources to influence policy decision-making processes, planning interventions, and wider public perceptions of accessibility and equity in urban environments. This study shows why and how we must think critically about the data we use when measuring urban accessibility. Future research could enhance this by modeling park entrances (for which consistent data simply do not exist currently) and open hours (to understand spatiotemporal accessibility). As case studies, we focused here on two cities with different sizes, histories, demographics, and economic conditions, but future broader sets of case studies could further help generalize these findings across more urban environments. Finally, not all data sources are transparent regarding provenance and representation. For example, Esri provided no information on the data collection process underlying their parks dataset, but noted it was sourced from TomTom, a for-profit navigation technology company. We asked Esri for more information, but they directed us to TomTom, who, in turn, was unwilling to provide any further information. The lack of transparency itself makes it difficult to understand and reason with the limitations and representational challenges of a dataset.

\section{Conclusion}

Our study extends the literature on accessibility and the lack of consensus on how to measure it, particularly to parks. We argue that biases in park accessibility analyses arise not only from the chosen method of measurement but also from the data source itself. Even when holding the impedances (e.g., network distance) constant, different data sources produce divergent analytical results. We argue that analysts must critically engage with data sources through precise conceptual definitions, theoretically-infused selection, triangulation of multiple sources, and careful validation to understand if their data truly represent their objects of interest.

In conclusion, this study demonstrates the importance of data source selection in urban accessibility measurements. We call for a more nuanced understanding of how source data---apart from calculation---shapes our understanding of urban landscapes. Through careful, theoretically-infused selection, filtering, and transformation of source data, we can achieve more-accurate analyses of urban accessibility.

\IfFileExists{\jobname.ent}{\theendnotes}{}

\setlength{\bibsep}{0.00cm plus 0.05cm} 
\bibliographystyle{apalike}
\bibliography{references}

\end{document}